\documentclass[11pt]{article}
\usepackage{tikz}
\usetikzlibrary{snakes}
\usepackage[utf8]{inputenc} 
\usepackage{textcomp} 
\usepackage{flafter}  

\usepackage{amsmath,amssymb}  
\usepackage{bm}  
\usepackage{authblk}
\usepackage{memhfixc}  
\usepackage{listings}
\numberwithin{equation}{section}
\begin{document}
\title{Exact Green's function of the reversible ABCD reaction in two space dimensions}
\author{Thorsten Pr\"ustel} 
\author{Martin Meier-Schellersheim} 
\affil{Laboratory of Systems Biology\\National Institute of Allergy and Infectious Diseases\\National Institutes of Health}
\maketitle
\let\oldthefootnote\thefootnote 
\renewcommand{\thefootnote}{\fnsymbol{footnote}} 
\footnotetext[1]{Email: prustelt@niaid.nih.gov, mms@niaid.nih.gov} 
\let\thefootnote\oldthefootnote 
\abstract
{We derive an exact expression for the Green's functions in the time domain of the reversible diffusion-influenced ABCD reaction $A+B\leftrightarrow C+D$ in two space dimensions. Furthermore, we calculate the corresponding survival and reaction probabilities. The obtained expressions should prove useful for the study of reversible membrane-bound reactions in cell biology and can serve as a useful ingredient of enhanced stochastic particle-based simulation algorithms.}
\section{Introduction}
The reversible so-called ABCD reaction refers to the following reaction schemes \cite{Popov:2002}
\begin{eqnarray}
&& A+B \overset{\kappa_{1}}{\underset{\kappa_{2}}{\rightleftharpoons}} C+D, \\
&& A+B \overset{\kappa_{1}}{\underset{\kappa_{2}}{\rightleftharpoons}} C+B. 
\end{eqnarray} 
In particular, it represents the enzymatic reaction
\begin{equation}\label{MMreaction}
E+S \rightleftharpoons E+P,
\end{equation} 
where the enzyme $E$ converts a substrate $S$ into a product. Note that the reaction of Eq.~\eqref{MMreaction} is the limiting case of 
\begin{equation}\label{MMreaction}
E+S \rightleftharpoons ES \rightleftharpoons E+P
\end{equation} 
under the assumption that the concentration of the intermediate complex does not change on the timescale of the association/dissociation reactions.

Exact analytical expressions for the Green's functions (GF) of the reversible ABCD reaction for an isolated pair have been derived for 1, 2 and 3 dimensions (3D) in the Laplace domain
and for $1$ and $3$ dimensions in the time domain \cite{Popov:2002}. However, an exact solution valid for all times in 2D is still lacking despite both its broad applicability to membrane phenomena
and its conceptual importance. 
In fact, in the context of cell biological applications, the 2D case provides the basis for a better understanding of processes such as signal-induced
inhomogeneities and receptor clustering on cell membranes \cite{bethani:2010}, where diffusion constants that govern the lateral motion of receptors are typically small.
Furthermore, from a conceptual point of view, diffusion in two dimensions is special, because 2D is the critical dimension with regard to recurrence and transience of random walks \cite{Toussaint_Wilczek_1983}.

Moreover, GFs for an isolated pair play an important role in a number of particle-based stochastic simulation algorithms, because any biochemical reaction network may be thought of as composed of unimolecular and bimolecular reactions. In this context, analytical expressions for GFs describing an isolated pair can be employed to enhance the efficiency and accuracy of Brownian dynamics simulations \cite{Edelstein:1993, kimYangShin:1999, Popov:2003, vanZon:2005p340}. 
\section{Theory}
In this section we will closely follow Ref.~\cite{Popov:2002}. 
The two different possible states of the ABCD reaction require considering two different probability density functions:  $p_{1}(r_{1}, t\vert \xi_{1})$ yields the probability to find an $A,B$ pair at a distance equal to $r_{1}$ at time $t$, given that it was initially separated by a distance equal to $\xi_{1}$, and $p_{2}(r_{2}, t\vert \xi_{1})$ provides the corresponding probability for the $C,D$ pair, given that initially an $A, B$ pair was separated by a distance equal to $\xi_{1}$. The relative diffusion constants in the two states are denoted by $D_{1}$ and $D_{2}$, respectively. When the distance equals the reaction distance $a_{1} (a_{2})$, the pair $A,B$ ($C,D$) may undergo a reaction
and be converted to $C,D$ ($A,B$). The transitions from one state to the other are characterized by the intrinsic rate constants $\kappa_{1}$ and $\kappa_{2}$, respectively.  
The time evolution of the probability densities is governed by the two equations \cite{Popov:2002}
\begin{eqnarray}\label{eqMotion}
&&\frac{\partial p_{i}(r_{i}, t\vert \xi_{1})}{\partial t} =D_{i}\bigg(\frac{\partial^{2}}{\partial r_{i}^{2}}+\frac{1}{r_{i}}\frac{\partial}{\partial r_{i}}\bigg)p_{i}(r_{i}, t\vert \xi_{1}) - \kappa_{i}\frac{\delta(r_{i} - a_{i})}{2\pi r_{i}}p_{i}(r_{i}, t\vert \xi_{1})\nonumber\\
&& + \kappa_{j}\frac{\delta(r_{i} - a_{i})}{2\pi r_{i}}p_{j}(a_{j}, t\vert \xi_{1}), \quad\text{where $i,j=1,2$ and $ i\neq j$.}
\end{eqnarray}    
The Eqs.~\eqref{eqMotion} have to be supplemented by an appropriate initial condition. As already mentioned, we assume, without loss of generality, that the initial state at time $t=0$ is given by an $A,B$ pair separated by a distance equal to $\xi_{1}$. Hence, the initial conditions read
\begin{eqnarray}
p_{1}(r_{1}, t=0\vert \xi_{1}) &=& \frac{\delta(r_{1}-\xi_{1})}{2\pi r_{1}}, \label{init1}\\
p_{2}(r_{2}, t=0\vert \xi_{1}) &=& 0. \label{init2}
\end{eqnarray} 
In the following we are interested in the GF $p_{1}(r_{1}, t\vert \xi_{1})$ and $p_{2}(r_{2}, t\vert \xi_{1})$ that satisfy the initial value problem posed by Eqs.~\eqref{eqMotion}, \eqref{init1}, \eqref{init2}.

Employing the Laplace transform 
\begin{equation}
\tilde{p}_{i}(r_{i}, s\vert \xi_{1}) = \int^{\infty}_{0}e^{-st}p_{i}(r_{i}, t\vert \xi_{1})dt, \quad i=1,2,
\end{equation}
Popov and Agmon \cite{Popov:2002} obtained explicit expressions for the GF in the Laplace domain
\begin{eqnarray}
\tilde{p}_{1}(r_{1}, s\vert \xi_{1}) &=& \tilde{G}_{1,\text{ref}}(r_{1}, s\vert \xi_{1})\nonumber\\
&&-\frac{\kappa_{1} \tilde{G}_{1,\text{ref}}(r_{1}, s\vert a_{1})\tilde{G}_{1,\text{ref}}(a_{1}, s\vert \xi_{1})}{1+\kappa_{1}\tilde{G}_{1,\text{ref}}(a_{1}, s\vert a_{1})+\kappa_{2}\tilde{G}_{2,\text{ref}}(a_{2}, s\vert a_{2})},\qquad\quad\label{laplaceGF1}\\
\tilde{p}_{2}(r_{2}, s\vert \xi_{1}) &=& \frac{\kappa_{1} \tilde{G}_{2,\text{ref}}(r_{2}, s\vert a_{2})\tilde{G}_{1,\text{ref}}(a_{1}, s\vert \xi_{1})}{1+\kappa_{1}\tilde{G}_{1,\text{ref}}(a_{1}, s\vert a_{1})+\kappa_{2}\tilde{G}_{2,\text{ref}}(a_{2}, s\vert a_{2})},\label{laplaceGF2}\
\end{eqnarray}
where $\tilde{G}_{i,\text{ref}}(r, s\vert \xi)$ denotes the Laplace transform of the non-reactive GF that satisfies at the encounter distance a reflective boundary condition
\begin{equation}
\frac{\partial G_{i,\text{ref}}(r, s\vert \xi)}{\partial r}\bigg\vert_{r = a_{i}} = 0, \quad i=1,2.
\end{equation}
Clearly, the probability density of a non-reactive isolated pair has to satisfy
\begin{equation}
2\pi\int^{\infty}_{a_{i}} \tilde{G}_{i,\text{ref}}(r_{i}, s\vert a_{i}) r_{i} dr_{i} = \frac{1}{s}, \quad i=1,2.
\end{equation}
Due to this property, the survival or reaction probabilities 
\begin{equation}
\tilde{S}_{i}(s\vert \xi_{1}) = 2\pi\int^{\infty}_{a_{i}} \tilde{p}_{i}(r_{i}, s\vert \xi_{i}) r_{i} dr_{i} 
\end{equation}
can easily be obtained from Eqs.~\eqref{laplaceGF1},~\eqref{laplaceGF2} \cite{Popov:2002}
\begin{eqnarray}
s\tilde{S}_{1}( s\vert \xi_{1}) &=& 1 -\frac{\kappa_{1} \tilde{G}_{1,\text{ref}}(a_{1}, s\vert \xi_{1})}{1+\kappa_{1}\tilde{G}_{1,\text{ref}}(a_{1}, s\vert a_{1})+\kappa_{2}\tilde{G}_{2,\text{ref}}(a_{2}, s\vert a_{2})},\qquad\label{laplaceS1}\\ 
s\tilde{S}_{2}(s\vert \xi_{1}) &=& \frac{\kappa_{1}\tilde{G}_{1,\text{ref}}(a_{1}, s\vert \xi_{1})}{1+\kappa_{1}\tilde{G}_{1,\text{ref}}(a_{1}, s\vert a_{1})+\kappa_{2}\tilde{G}_{2,\text{ref}}(a_{2}, s\vert a_{2})}.\label{laplaceS2}
\end{eqnarray}
It follows immediately from Eqs.~\eqref{laplaceS1},~\eqref{laplaceS2}  that $\tilde{S}_{1}( s\vert \xi_{1}) + \tilde{S}_{2}( s\vert \xi_{1}) = 1/s$, which translates to
\begin{equation}\label{conservationProb}
S_{1}( t\vert \xi_{1}) + S_{2}( t\vert \xi_{1}) = 1,
\end{equation}
in the time domain, as it should be.
\section{Explicit form of the Green's functions and survival probabilities}
Eqs.~\eqref{laplaceGF1},~\eqref{laplaceGF2} permit obtaining $\tilde{p}_{i}(r_{i}, s\vert \xi_{1})$ once the explicit form of the non-reactive GF in the Laplace domain is known.
In 2D, it is given by \cite{Popov:2002}
\begin{equation}\label{laplaceGFref}
\tilde{G}_{i,\text{ref}}(r_{i}, s\vert \xi_{i}) = \tilde{G}_{i, \text{free}}(r_{i}, s\vert \xi_{i})+\frac{1}{2\pi D_{i}} K_{0}(q_{i}r_{i})K_{0}(q_{i}\xi_{i})\frac{I_{1}(q_{i}a_{i})}{K_{1}(q_{i}a_{i})}.
\end{equation}
Here, $I_{n}, K_{n}$ refer to the modified Bessel function of first and second kind, respectively \cite[Sect.~9.6]{abramowitz1964handbook} and $\tilde{G}_{i,\text{free}}(r, s\vert \xi)$ denotes the free-space GF \cite[Ch. 14.8, Eq. (2)]{carslaw1986conduction}
\begin{equation}\label{laplaceFree}
\tilde{G}_{i\text{free}}(r_{i}, s \vert \xi_{i}) = \frac{1}{2\pi D_{i}}
\biggl\{\begin{array}{lr}
 I_{0}(q_{i}\xi_{i}) K_{0}(q_{i}r_{i}),&\text{$ r_{i}  >  \xi_{i} $} \\
 I_{0}(q_{i}r_{i}) K_{0}(q_{i}\xi_{i}), &\text{$r_{i}  <  \xi_{i}$}  
\end{array}
\end{equation}
and we define $q_{i}=\sqrt{s/D_{i}}$ for $i=1,2$.
Combining Eqs.~\eqref{laplaceGF2},~\eqref{laplaceS2} with Eqs.~\eqref{laplaceGFref},~\eqref{laplaceFree} one arrives at
\begin{eqnarray}\label{inversionFormula}
\tilde{p}_{2}(r_{2}, t \vert \xi_{1}) &=& \frac{h_{1}}{2\pi D_{2}a_{2}} \frac{K_{0}(q_{2}r_{2})K_{0}(q_{1}\xi_{1})}{q_{2}K_{1}(q_{2}r_{2})}\frac{1}{\mathcal{D}}, \\
\tilde{S}_{2}(t \vert \xi_{1}) &=&  \frac{h_{1}}{s} \frac{K_{0}(q_{1}\xi_{1})}{\mathcal{D}},
\end{eqnarray}
where the denumerator $\mathcal{D}$ is given by
\begin{equation}\label{denumerator}
\mathcal{D} = q_{1}K_{1}(q_{1}a_{1})\bigg[1 +h_{2}\frac{K_{0}(q_{2}a_{2})}{q_{2}K_{1}(q_{2}a_{2})}\bigg] + h_{1}K_{0}(q_{1}a_{1}).
\end{equation}
Here, we introduced
\begin{equation}
h_{i} = \kappa_{i}/(2\pi D_{i}a_{i}), \quad i =1,2.
\end{equation}

In the following we employ the inversion theorem for the Laplace transformation to obtain the corresponding GF and survival probabilities in the time domain according to
\begin{eqnarray}\label{inversionFormula}
p_{i}(r_{i}, t \vert \xi_{1}) &=& \frac{1}{2\pi i} \int^{\gamma+i\infty}_{\gamma-i\infty} e^{st}\,\tilde{p}_{i}(r_{i}, s\vert \xi_{1} )ds,\\
S_{i}(t \vert \xi_{1}) &=& \frac{1}{2\pi i} \int^{\gamma+i\infty}_{\gamma-i\infty} e^{st}\,\tilde{S}_{i}(s\vert \xi_{1} )ds.
\end{eqnarray}

\begin{figure}
\includegraphics[scale=0.29]{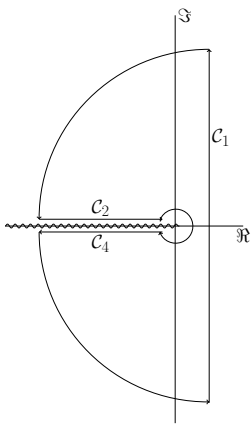}
\caption{Integration contour used  in Eq.~(\ref{cauchy}).}\label{fig:contour}
\end{figure}

It is convenient to consider $p_{2}(r_{2}, t \vert \xi_{1})$ first. 
We proceed analogously to the calculations presented in Ref. \cite{TPMMS_2012JCP}, where the GF of the reversible diffusion-influenced reaction $A+B\rightleftharpoons C$ for an isolated pair  in two space dimensions was obtained. 
In particular, to calculate the Bromwich contour integral, we note that $\tilde{p}_{2}(r_{2}, t \vert \xi_{1})$ is multi-valued and has a branch point at $s=0$. Therefore, we consider the contour of Fig.~\ref{fig:contour} with a branch cut along the negative real axis, cp. \cite[Ch. 12.3, Fig. 40]{carslaw1986conduction}.  In addition, we note that the integrand has no poles within and on the contour  \cite{erdelyiKermack:1945} and the contribution from the small circle around the origin vanishes. Thus, the only contributions come from the integration above and below the branch cut 
\begin{eqnarray}\label{cauchy}
&&\int^{\gamma+i\infty}_{\gamma-i\infty} e^{st}\,\tilde{p}_{2}(r_{2}, s\vert \xi_{1} )ds = \nonumber\\
&& - \int_{\mathcal{C}_{2}} e^{st}\,\tilde{p}_{2}(r_{2}, s\vert \xi_{1} )ds - \int_{\mathcal{C}_{4}} e^{st}\,\tilde{p}_{2}(r_{2}, s\vert \xi_{1} )ds.
\end{eqnarray}
To calculate the integrals $\int_{\mathcal{C}_{2}}, \int_{\mathcal{C}_{4}}$, 
we choose, without loss of generality,
$
s = D_{1} x^{2} e^{i \pi }
$
and use \cite[Append.~3, Eqs.~(25), (26))]{carslaw1986conduction}
\begin{eqnarray}
I_{n}(xe^{\pm \pi i/2}) &=& e^{\pm n\pi i/2} J_{n}(x), \\
K_{n}(xe^{\pm \pi i/2}) &=& \pm\frac{1}{2}\pi i e^{\mp n\pi i/2} [-J_{n}(x) \pm i Y_{n}(x)].
\end{eqnarray}
$J_{n}(x), Y_{n}(x)$ denote the Bessel functions of first and second kind, respectively \cite[Sect.~9.1]{abramowitz1964handbook}.
It follows that
\begin{eqnarray}
&&\int_{\mathcal{C}_{2}}e^{st}\,\tilde{p}_{2}(r_{2}, s\vert \xi_{1} )ds  =  \frac{h_{1}}{\pi a_{2}}\sqrt{\frac{D_{1}}{D_{2}}}\int^{\infty}_{0}e^{-D_{1}x^{2}t} \times  \qquad\qquad\qquad\nonumber \\
&& \frac{[\Omega(r_{2}\sqrt{D_{1}/D_{2}}, \xi_{1}) - i\Pi(r_{2}\sqrt{D_{1}/D_{2}}, \xi_{1})][-Y_{1}(x\phi)+iJ_{1}(x\phi)][\beta-i\alpha]}{[\alpha^{2} + \beta^{2}][J^{2}_{1}(x\phi)+Y^{2}_{1}(x\phi)]} dx.\qquad\qquad
\end{eqnarray}
Here, we introduced a fair amount of new notation.
First, we have
\begin{eqnarray}
\alpha &=& \alpha_{\text{rad}} + \alpha_{b}, \\
\beta &=& \beta_{\text{rad}} + \beta_{b}, 
\end{eqnarray}
where $\alpha_{\text{rad}}, \beta_{\text{rad}}$ refer to the functions that appear in the expression of the GF that describes irreversible association of an isolated pair in 2D, i.e.  a GF that satisfies
a radiation boundary condition at contact $a_{1}$ \cite[Chap. 14.8, Eqs.~(12), (13)]{carslaw1986conduction}
\begin{equation}\label{GFrad}
p_{\text{rad}}(r, t\vert \xi )=\frac{1}{2\pi}\int^{\infty}_{0}e^{-D_{1}tx^{2}}T_{\text{rad}}(x, r)T_{\text{rad}}(x, \xi)\,x\,dx,
\end{equation}
where 
\begin{equation}\label{defTrad}
T_{\text{rad}}(x, r) = \frac{J_{0}(rx)\beta_{\text{rad}} - Y_{0}(rx)\alpha_{\text{rad}}}{[\alpha^{2}_{\text{rad}}+\beta^{2}_{\text{rad}}]^{1/2}},
\end{equation}
and
\begin{eqnarray}
\alpha_{\text{rad}} &=& xJ_{1}(xa_{1}) + h_{1}J_{0}(xa_{1}), \\
\beta_{\text{rad}} &=& xY_{1}(xa_{1}) + h_{1}Y_{0}(xa_{1}).
\end{eqnarray}
Furthermore, we defined
\begin{eqnarray}
\alpha_{b} &=& \sigma Y_{1}(xa_{1}) + \rho J_{1}(xa_{1}), \\
\beta_{b} &=& \rho Y_{1}(xa_{1}) - \sigma J_{1}(xa_{1}), 
\end{eqnarray}
where
\begin{eqnarray}
\rho &=& \psi (Y_{0}(x\phi)Y_{1}(x\phi) + J_{0}(x\phi )J_{1}(x\phi)), \\
\sigma &=& -\psi \frac{2}{\pi x\phi}, 
\end{eqnarray}
and
\begin{eqnarray}
\psi = h_{2}\sqrt{\frac{D_{2}}{D_{1}}}\frac{1}{[J_{1}(x\phi)^{2} + Y_{1}(x\phi)^{2}]}, \quad\phi = \sqrt{\frac{D_{1}}{D_{2}}}a_{2}.
\end{eqnarray}
Finally, we introduced
\begin{eqnarray}
\Omega(r, \xi) &=& J_{0}(xr)J_{0}(x\xi) - Y_{0}(xr)Y_{0}(x\xi), \\
\Pi(r, \xi) &=& Y_{0}(xr)J_{0}(x\xi) + J_{0}(xr)Y_{0}(x\xi). 
\end{eqnarray}

Next, we consider the integral along the contour $\mathcal{C}_{4}$ below the branch cut. It can be evaluated by choosing $s = D_{1}x^{2}e^{-i\pi}$ and after a calculation analogous to the evaluation of $\int_{\mathcal{C}_{2}}$ we obtain 
\begin{equation}
\int_{\mathcal{C}_{2}}e^{ps}\,\tilde{p}_{2}(r_{2}, s\vert \xi_{1})ds=-\bigg(\int_{\mathcal{C}_{4}}e^{st}\,\tilde{p}_{2}(r_{2}, s\vert \xi_{1} )ds\bigg)^{\ast},
\end{equation}
where $\ast$ means complex conjugation.
Thus, the exact Green's function in the time domain is given by
\begin{eqnarray}\label{GF2}
&&p_{2}(r_{2}, t \vert \xi_{1}) = -\frac{1}{\pi} \Im\bigg(\int_{\mathcal{C}_{2}}e^{st}\,\tilde{p}_{2}(r_{2}, s\vert \xi_{1} )ds\bigg)\nonumber \\
&&= -\frac{h_{1}}{\pi^{2} a_{2}}\sqrt{\frac{D_{1}}{D_{2}}}\int^{\infty}_{0}e^{-D_{1}x^{2}t} \times\nonumber\\
&&\bigg\lbrace\frac{\alpha[Y_{1}(x\phi)\Omega(r_{2}\sqrt{D_{1}/D_{2}}, \xi_{1}) - J_{1}(x\phi)\Pi(r_{2}\sqrt{D_{1}/D_{2}}, \xi_{1})]}{[\alpha^{2} + \beta^{2}][J^{2}_{1}(x\phi)+Y^{2}_{1}(x\phi)]} \nonumber \\
&&+ \frac{\beta[J_{1}(x\phi)\Omega(r_{2}\sqrt{D_{1}/D_{2}}, \xi_{1}) + Y_{1}(x\phi)\Pi(r_{2}\sqrt{D_{1}/D_{2}}, \xi_{1})]}{[\alpha^{2} + \beta^{2}][J^{2}_{1}(x\phi)+Y^{2}_{1}(x\phi)]}\bigg\rbrace dx.\qquad \qquad
\end{eqnarray}

To obtain the corresponding reaction probability $S_{2}(t\vert\xi_{1})$, we can proceed analogously to the calculation of the GF $p_{2}(r_{2}, t \vert \xi_{1})$, in particular, we can employ the same integration contour.
There is one important difference, though. In the case of the survival probability, the contribution from the integration along the small circle around the origin does not vanish. To calculate this contribution we use $s= D_{1}x^{2}e^{i\varphi}$ and integrate from $-\pi$ to $\pi$. Using Eq.~\eqref{laplaceS2},~\eqref{denumerator} and the small argument expansions of the modified Bessel functions \cite[Eqs.~(9.6.7)-(9.6.9)]{abramowitz1964handbook}, we arrive
at
\begin{eqnarray}
&& 2\pi i \,\, \text{contribution from the integral around the origin} \nonumber\\
&& = -i\lim_{x\rightarrow 0}\int^{-\pi}_{\pi} \frac{h_{1}K_{0}(q_{1}\xi_{1})}{\mathcal{D}}d\varphi =  -i \lim_{x\rightarrow 0}\int^{-\pi}_{\pi} \frac{h_{1}\ln x}{h_{1}\ln x + 
h_{2}a_{2}/a_{1} \ln x}d\varphi  \nonumber \\
&& =2\pi i \frac{h_{1}a_{1}}{h_{1}a_{1} + h_{2}a_{2}}.
\end{eqnarray}
Taking into account the contributions from the integral above and below the branch cut and around the origin, we finally obtain 
\begin{eqnarray}\label{S2}
S_{2}(t \vert \xi_{1}) &=& \frac{h_{1}a_{1}}{h_{1}a_{1} + h_{2}a_{2}} \nonumber\\
&&- \frac{2h_{1}}{\pi}\int^{\infty}_{0}e^{-D_{1}tx^{2}} \frac{\alpha Y_{0}(x\xi_{1}) - \beta J_{0}(x\xi_{1})}{\alpha^{2}+\beta^{2}} \frac{dx}{x}.
\end{eqnarray}
In particular, we get for the steady-state values of $S_{2}(t\vert \xi_{1})$ and of $S_{1}(t\vert \xi_{1})$ (due to Eq.~\eqref{conservationProb}) 
\begin{eqnarray}
\lim_{t\rightarrow \infty}S_{1}(t\vert \xi_{1}) &=& \frac{h_{2}a_{2}}{h_{1}a_{1} + h_{2}a_{2}},\\
\lim_{t\rightarrow \infty}S_{2}(t\vert \xi_{1}) &=& \frac{h_{1}a_{1}}{h_{1}a_{1} + h_{2}a_{2}},
\end{eqnarray}
in agreement with Ref.~\cite{Popov:2002}.
Also, the explicit form of $S_{1}(t\vert \xi_{1})$ follows immediately from Eqs.~\eqref{S2},~\eqref{conservationProb}.

It remains to calculate the GF $p_{1}(r_{1}, t\vert \xi_{1})$. We know from Eq.~\eqref{laplaceGF1} that it can be written as
\begin{equation}\label{p1Result}
p_{1}(r_{1}, t\vert \xi_{1}) = G_{1,\text{ref}}(r_{1}, t\vert \xi_{1}) + p_{1\text{bc}}(r_{1}, t\vert \xi_{1}),
\end{equation}  
where the explicit form of $G_{1,\text{ref}}(r_{1}, t\vert \xi_{1})$ is well-known and can be recovered from Eq.~\eqref{GFrad} in the limit $h_{1}\rightarrow 0$.
$p_{1\text{bc}}(r_{1}, t\vert \xi_{1})$ can easily be obtained from Eq.~\eqref{GF2}, exploiting the similarity of $\tilde{p}_{1\text{bc}}(r_{1}, s\vert \xi_{1})$ and $\tilde{p}_{2}(r_{2}, s\vert \xi_{1})$, cp.
Eqs.~\eqref{laplaceGF1}, \eqref{laplaceGF2}. Thus, we obtain
\begin{eqnarray}\label{GF1}
&&p_{1bc}(r_{1}, t \vert \xi_{1}) = \frac{h_{1}}{\pi^{2} a_{1}}\int^{\infty}_{0}e^{-D_{1}x^{2}t} \times\nonumber\\
&& \bigg\lbrace\frac{\alpha[Y_{1}(xa_{1})\Omega(r_{1}, \xi_{1}) - J_{1}(xa_{1})\Pi(r_{1}, \xi_{1})]}{[\alpha^{2} + \beta^{2}][J^{2}_{1}(xa_{1})+Y^{2}_{1}(xa_{1})]} \nonumber\\
&&+ \frac{\beta[J_{1}(xa_{1})\Omega(r_{1}, \xi_{1}) + Y_{1}(xa_{1})\Pi(r_{1}, \xi_{1})]}{[\alpha^{2} + \beta^{2}][J^{2}_{1}(xa_{1})+Y^{2}_{1}(xa_{1})]}\bigg\rbrace   dx.\qquad \qquad
\end{eqnarray}
Note that in the limit $\kappa_{2}\rightarrow 0$ one recovers from Eqs.~\eqref{p1Result},\eqref{GF1} the known GF that satisfies a radiation BC Eq.~\eqref{GFrad} and describes the irreversible association $A+B\rightarrow C$.
\subsection*{Acknowledgments}
This research was supported by the Intramural Research Program of the NIH, National Institute of Allergy and Infectious Diseases. 
\newpage
\bibliographystyle{plain}

\begin{thebibliography}{10}
\bibitem{abramowitz1964handbook}
M.~Abramowitz and I.A. Stegun.
\newblock {\em Handbook of Mathematical Functions with Formulas, Graphs, and
  Mathematical Tables}.
\newblock Dover, New York, 1965.

\bibitem{bethani:2010}
I.~Bethani, S.S. Skanland, I.~Dikic, and A.~Acker-Palmer.
\newblock {\em EMBO J.}, 29:2677, 2010.

\bibitem{carslaw1986conduction}
H.S. Carslaw and J.C. Jaeger.
\newblock {\em Conduction of Heat in Solids}.
\newblock Clarendon Press, New York, 1986.

\bibitem{Edelstein:1993}
A.L. Edelstein and N.~Agmon.
\newblock {\em J. Chem. Phys.}, 99:5396, 1993.

\bibitem{erdelyiKermack:1945}
A.~Erdelyi and W.O. Kermack.
\newblock {\em Proc. Camb. Phil. Soc.}, 41:74, 1945.

\bibitem{kimYangShin:1999}
H.~Kim, M.~Yang, and K.J. Shin.
\newblock {\em J. Chem. Phys.}, 111:1068, 1999.

\bibitem{Popov:2002}
A.V. Popov and N.~Agmon.
\newblock {\em J. Chem. Phys.}, 117:5770, 2002.

\bibitem{Popov:2003}
A.V. Popov and N.~Agmon.
\newblock {\em J. Chem. Phys.}, 118:11057, 2003.

\bibitem{TPMMS_2012JCP}
T.~Pr\"ustel and M.~Meier-Schellersheim.
\newblock {\em J. Chem. Phys.}, 137:054104, 2012.

\bibitem{Toussaint_Wilczek_1983}
D.~Toussaint and F.~Wilczek.
\newblock {\em J. Chem. Phys.}, 78:2642, 1983.

\bibitem{vanZon:2005p340}
J.S. van Zon and P.R. ten Wolde.
\newblock {\em Phys. Rev. Lett.}, 94:128103, 2005.

\end{thebibliography}

\end{document}